\documentclass[pdflatex,sn-mathphys-num]{sn-jnl}
\usepackage{adjustbox}
\usepackage{xcolor}
\usepackage{float}
\usepackage[utf8]{inputenc}
\usepackage{tikz}
\usepackage{chemfig}
\usepackage{booktabs} 
\usetikzlibrary{shapes.geometric, arrows.meta, positioning}
\usepackage{longtable}
\usetikzlibrary{shapes.geometric, arrows.meta}
\usepackage{graphicx}%
\usepackage{multirow}%
\usepackage{amsmath,amssymb,amsfonts}%
\usepackage{amsthm}%
\usepackage{mathrsfs}%
\usepackage[title]{appendix}%
\usepackage{xcolor}%
\usepackage{textcomp}%
\usepackage{manyfoot}%
\usepackage{booktabs}%
\usepackage{algorithm}%
\usepackage{algorithmicx}%
\usepackage{algpseudocode}%
\usepackage{listings}%
\theoremstyle{thmstyleone}%

\theoremstyle{thmstyletwo}%
\theoremstyle{thmstylethree}%
\usepackage{siunitx}
\begin{document}

\title[A Comparative Study of Exponential Sum-Connectivity and Product-Connectivity Gourava Indices for Benzenoid Hydrocarbons]{A Comparative Study of Exponential Sum-Connectivity and Product-Connectivity Gourava Indices for Benzenoid Hydrocarbons}

\author*[1]{\fnm{H. M. Nagesh}}\email{nageshhm@pes.edu}
\author[2]{\fnm{B. Azghar Pasha}}\email{azgharpasha@msrit.edu}
\author[3]{\fnm{U. Vijaya Chandra Kumar}}\email{uvijaychandra.kumar@reva.edu.in}
\author[4]{\fnm{Narahari N}}\email{narahari\_nittur@yahoo.com}

\affil*[1]{\orgdiv{Department of Science and Humanities}, \orgname{PES University}, \orgaddress{\city{Bangalore}, \state{Karnataka}, \country{India}}}

\affil[2]{\orgdiv{Department of Mathematics}, \orgname{M. S. Ramaiah Institute of Technology}, \orgaddress{\city{Bangalore}, \state{Karnataka}, \country{India}}}

\affil[3]{\orgdiv{Department of Mathematics, School of Applied Sciences}, \orgname{REVA University}, \orgaddress{\city{Bangalore}, \state{Karnataka}, \country{India}}}

\affil[4]{\orgdiv{Department of Mathematics, University College of Science}, \orgname{Tumkur University}, \orgaddress{\city{Tumakuru}, \state{Karnataka}, \country{India}}}

\abstract{In this work, the exponential sum-connectivity Gourava index ($e^{SGO}$) and the exponential product-connectivity Gourava index ($e^{PGO}$) are computed and comparatively analyzed for benzenoid hydrocarbons. Our results demonstrate that these descriptors exhibit a strong mutual correlation and provide enhanced sensitivity in modeling the structural characteristics of molecular graphs. Regression analysis reveals that both indices are exceptionally reliable predictors of $\pi$-electronic energies, achieving correlation coefficients exceeding $0.999$. Notably, a comparative assessment indicates that the exponential product-connectivity variant offers a slightly superior fit, as its coefficients align more precisely with optimal least-squares results. These findings confirm that both exponential Gourava-based indices provide a robust framework for characterizing electronic properties, with the product-connectivity version showing particular promise for high-precision QSPR studies in benzenoid systems.}


\maketitle
\section{Introduction}
\label{sec1}
In chemical graph theory, a molecule's structure is modelled as a graph $G = (V, E)$, where vertices $V$ represent individual atoms and edges $E$ represent the chemical bonds connecting them. To quantify these structures, researchers utilize topological indices. A topological index is a unique numerical value derived from the graph that remains constant regardless of how the graph is drawn (graph isomorphism). These descriptors are essential tools in Quantitative Structure-Property Relationship (QSPR) and Quantitative Structure-Activity Relationship (QSAR) studies, allowing scientists to predict physical and chemical properties — such as boiling points and molecular stability — solely from the graph's geometry. Among these descriptors, indices based on vertex degrees (the number of bonds attached to an atom) are particularly prominent. Recently, V. R. Kulli introduced the sum-connectivity Gourava index and the product-connectivity Gourava index \cite{1,2}. These modern descriptors offer a more detailed analysis of molecular branching and connectivity by mathematically combining the sums and products of vertex degrees.

For any edge $uv \in G$ in the edge set $E(G)$, let $d_G(u)$ and $d_G(v)$ denote the degrees of the adjacent vertices $u$ and $v$, respectively. Then, the sum-connectivity Gourava index and the product-connectivity Gourava index are defined as follows:

\begin{equation*}
SGO=SGO(G)= \sum_{uv \in E(G)} \frac{1}{\sqrt{d_{G}(u) + d_{G}(v) + d_{G}(u)d_{G}(v)}}
\end{equation*}

\begin{equation*}
PGO=PGO(G)= \sum_{uv \in E(G)} \frac{1}{\sqrt{(d_{G}(u) + d_{G}(v))(d_{G}(u)d_{G}(v))}}
\end{equation*}

The practical utility and predictive power of the Gourava indices within the framework of QSPR analysis were recently established in \cite{3}. The discriminating ability of topological indices plays a crucial role in the analysis and comparison of molecular descriptors \cite{4}. Motivated by this consideration, the exponential form of a vertex-degree-based topological index was introduced in \cite{5}. The investigation conducted in \cite{6} validates the efficacy of exponential topological indices in characterizing various physicochemical properties of octane isomers through QSPR modeling. In the present work, we explore the applicability of the exponential sum-connectivity and product-connectivity Gourava indices to the structural modelling of benzenoid hydrocarbons. 

Benzenoid hydrocarbons constitute a fundamental class of polycyclic aromatic hydrocarbons (PAHs) characterized by the fusion of hexagonal benzene rings. Due to their unique electronic stability and structural symmetry, these molecules serve as essential models in both organic chemistry and materials science. \newpage In chemical graph theory, benzenoids are represented as planar graphs where each internal face is a regular hexagon, providing a rich framework for topological analysis. Understanding their physicochemical behaviour is critical, as these compounds are widely encountered in environmental chemistry and industrial synthesis. 

We evaluate the predictive power of these indices by correlating them with $\pi$-electronic energies, demonstrating their effectiveness as reliable topological descriptors for aromatic molecular systems. Fig. 1 illustrates the different types of inlets occurring in benzenoid systems \cite{7}.

\begin{figure}[H]
    \centering
    \begingroup
    \setlength{\fboxsep}{0pt}
    \colorbox{black!15}{%
        \includegraphics[width=0.70\linewidth]{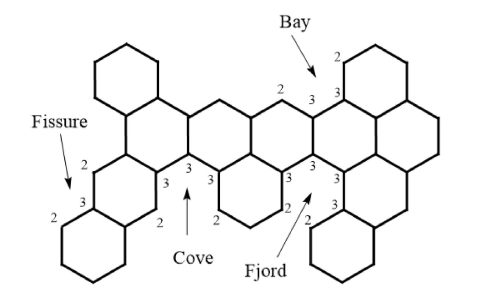}%
    }
    \endgroup
    \caption{Representation of bays, fjords, fissures, and coves in benzenoid systems.}
    \label{fig:placeholder}
\end{figure}

\subsection{Motivation and Novelty}
The motivation for this study arises from the established success of classical connectivity indices in characterizing polycyclic aromatic systems. Previous studies \cite{8} have demonstrated that the product-connectivity (Randi\'{c}) index and the sum-connectivity index exhibit strong correlations with the $\pi$-electronic energies of benzenoid hydrocarbons. In recent years, Gourava indices have emerged as versatile topological descriptors, offering a more precise mathematical framework for representing the structural complexity and vertex-degree distributions within molecular graphs. Furthermore, a notable advancement in this field involves the development of exponential variations of such indices, which aim to increase the sensitivity and discriminatory power of molecular descriptors.

The novelty of the present work lies in addressing a significant gap in the current literature. While standard Gourava-type measures have demonstrated strong predictive capabilities, a comprehensive evaluation of their exponential sum-connectivity and product-connectivity counterparts in the context of benzenoid hydrocarbons has not yet been established. Accordingly, this study provides the first rigorous comparative analysis to determine whether these exponential variations offer superior predictive accuracy for the $\pi$-electronic properties of benzenoid hydrocarbons, thereby extending the utility of Gourava-based descriptors in chemical graph theory.

\section{Exponential sum-connectivity and exponential product-connectivity Gourava indices of benzenoid hydrocarbons}
In this section, we find the exponential sum-connectivity and exponential product-connectivity Gourava indices of benzenoid hydrocarbons. The exponential sum-connectivity and exponential product-connectivity Gourava indices are defined as follows: 

\begin{equation}
e^{SGO}=e^{SGO}(G)= \sum_{uv \in E(G)} e^{\frac{1}{\sqrt{d_{G}(u) + d_{G}(v) + d_{G}(u)d_{G}(v)}}}
\end{equation}

\begin{equation}
e^{PGO}=e^{PGO}(G)= \sum_{uv \in E(G)} e^{\frac{1}{\sqrt{(d_{G}(u) + d_{G}(v))(d_{G}(u)d_{G}(v))}}}
\end{equation}

The structure of benzenoid hydrocarbons is most easily shown and studied using benzenoid graphs, which are mathematical drawings made of hexagonal rings. In benzenoid graphs, every vertex has a degree of either $2$ or $3$ \cite{9}, resulting in only three types of edges: $e_{22}$, $e_{23}$, and $e_{33}$. By introducing the number of these specific edges into the general definitions for exponential Gourava indices, we obtain simplified formulas for this class of molecules. The constant terms for these edges are calculated as follows:

\begin{table}[h!]
\centering
\caption{Calculated values of exponential Gourava indices for different edge types.}
\label{tab:edge_values}
\renewcommand{\arraystretch}{2.8} 
\begin{tabular}{|c|c|c|}
\hline
\textbf{Edge Type $(d_{G}(u), d_{G}(v))$} & \textbf{$e^{SGO}$} & \textbf{$e^{PGO}$} \\ \hline
(2, 2) & $e^{1/\sqrt{8}} \approx 1.4241$ & $e^{1/\sqrt{16}} \approx 1.2840$ \\ \hline
(2, 3) & $e^{1/\sqrt{11}} \approx 1.3519$ & $e^{1/\sqrt{30}} \approx 1.2003$ \\ \hline
(3, 3) & $e^{1/\sqrt{15}} \approx 1.2946$ & $e^{1/\sqrt{54}} \approx 1.1458$ \\ \hline
\end{tabular}
\end{table}

Substituting the calculated edge-weights and their corresponding counts ($e_{22}, e_{23}, e_{33}$) into (1) and (2) yields:
\begin{equation}
e^{SGO} = 1.4241 e_{22} + 1.3519 e_{23} + 1.2946 e_{33}
\end{equation}

\begin{equation}
e^{PGO} = 1.2840 e_{22} + 1.2003 e_{23} + 1.1458 e_{33} 
\end{equation}

Alternative expressions for (3) and (4) can be derived by considering the basic structural parameters of the graph. If a benzenoid system consists of $n$ vertices, $r$ inlets, and $h$ hexagons, the number of edges for each type is calculated as follows \cite{9}:

\begin{equation}
e_{22} = n - 2h - r + 2
\end{equation}
\begin{equation}
e_{23} = 2r
\end{equation}
\begin{equation}
e_{33} = 3h - r - 3
\end{equation}

The perimeter of these systems exhibits specific arrangements of vertex degrees called $inlets$, as established in the literature \cite{9,10}. These structural features are categorized into four distinct types: fissures ($f$), bays ($b$), coves ($c$), and fjords ($fj$). Each type is defined by its characteristic degree sequence along the boundary—specifically 232 for fissures, 2332 for bays, 23332 for coves, and 233332 for fjords (as shown in Figure 1). The total number of inlets, $r$, is calculated as the sum $r = f + b + c + fj$.

Substituting the edge counts from (5)-(7) into the definition for the exponential sum-connectivity and exponential product-connectivity Gourava indices leads to a concise formula expressed solely in terms of the number of vertices, inlets, and hexagons as follows:

\begin{align}
e^{SGO} &= 1.4241n + 1.0356h - 0.0149r - 1.0356 \tag{8} 
\end{align}
\begin{align}
e^{PGO} &= 1.2840n + 0.8694h -0.0292r - 0.8694 \tag{9}
\end{align}

Based on the mathematical relationships defined in (8) and (9), the exponential sum-connectivity and exponential product-connectivity Gourava indices were calculated for a sample of 30 benzenoid hydrocarbons. These results, alongside the corresponding $\pi$-electronic energy ($E_\pi$) values taken from \cite{8}, are systematically tabulated in Table 2.

\begin{table}[ht]
\centering
\caption{$\pi$-electronic energy ($E_{\pi}$), the exponential sum-connectivity index ($e^{SGO}$) and the exponential product-connectivity index ($e^{PGO}$) of 30 lower benzenoids.}
\begin{tabular}{clrrr}
\toprule
\textbf{Serial Number} & \textbf{Benzenoid Hydrocarbon} & $E_{\pi}$ & $e^{SGO}$ & $e^{PGO}$ \\
\midrule
1  & Benzene                  & 8.0000  & 8.5447  & 7.7042  \\
2  & Naphthalene              & 13.6832 & 15.2469 & 13.6511 \\
3  & Anthracene               & 19.3137 & 21.9491 & 19.5981 \\
4  & Phenanthrene             & 19.4483 & 21.9640 & 19.6273 \\
5  & Tetracene                & 24.9308 & 28.6513 & 25.5451 \\
6  & Benzo[c]phenanthrene     & 25.1875 & 28.6811 & 25.6035 \\
7  & Benzo[a]anthracene       & 25.1012 & 28.6662 & 25.5743 \\
8  & Chrysene                 & 25.1922 & 28.6811 & 25.6035 \\
9  & Triphenylene             & 25.2745 & 28.6961 & 25.6327 \\
10 & Pyrene                   & 22.5055 & 25.8329 & 23.0355 \\
11 & Pentacene                & 30.5440 & 35.3535 & 31.4921 \\
12 & Benzo[a]tetracene        & 30.7255 & 35.3684 & 31.5213 \\
13 & Dibenzo[a,h]anthracene   & 30.8805 & 35.3833 & 31.5505 \\
14 & Dibenzo[a,j]anthracene   & 30.8795 & 35.3833 & 31.5505 \\
15 & Pentaphene               & 30.7627 & 35.3684 & 31.5213 \\
16 & Benzo[g]chrysene         & 30.9990 & 35.4132 & 31.6089 \\
17 & Pentahelicene            & 30.9362 & 35.3982 & 31.5797 \\
18 & Benzo[c]chrysene         & 30.9386 & 35.3982 & 31.5797 \\
19 & Picene                   & 30.9432 & 35.3982 & 31.5797 \\
20 & Benzo[b]chrysene         & 30.8390 & 35.3833 & 31.5505 \\
21 & Dibenzo[a,c]anthracene   & 30.9418 & 35.3982 & 31.5797 \\
22 & Dibenzo[b,g]phenanthrene & 30.8336 & 35.3833 & 31.5505 \\
23 & Perylene                 & 28.2453 & 32.5649 & 29.0408 \\
24 & Benzo[e]pyrene           & 28.3361 & 32.5649 & 29.0408 \\
25 & Benzo[a]pyrene           & 28.2220 & 32.5500 & 29.0116 \\
26 & Hexahelicene             & 36.6814 & 42.1154 & 37.5559 \\
27 & Benzo[ghi]perylene       & 31.4251 & 36.4338 & 32.4490 \\
28 & Hexacene                 & 36.1557 & 42.0557 & 37.4391 \\
29 & Coronene                 & 34.5718 & 40.3027 & 35.8571 \\
30 & Ovalene                  & 46.4974 & 54.7725 & 48.6788 \\
\bottomrule
\end{tabular}
\end{table}

\newpage
\section{Comparative analysis of exponential sum-connectivity and exponential product-connectivity Gourava Indices}
To assess the statistical dependency between the exponential sum-connectivity Gourava index $e^{SGO}$ and the exponential product-connectivity Gourava index $e^{PGO}$, both measures were determined for the set of 30 benzenoid systems (Table 2). The linear correlation between these two topological descriptors is established through regression analysis. The resulting model, including the predictive equation and corresponding statistical validation parameters, is presented in Table 3.

\newpage
\begin{table}[ht]
\centering
\caption{Linear regression model and statistical validation parameters.}
\label{tab:regressiontable}
\begin{tabular}{ll}
\toprule
\textbf{Parameter} & \textbf{Value} \\
\midrule
Regression equation & $e^{PGO} = (0.8873 \pm 0.0006) e^{SGO} + (0.1415 \pm 0.0191)$ \\
Sample size ($N$)   & 30 \\
$S$ & 0.0259 \\
$R$   & 1.0000 \\
$S_{cv}$ & 0.0296 \\
$R_{cv}$ & 1.0000 \\
\bottomrule
\end{tabular}
\end{table}

Here, $R$ and $R_{cv}$ represent the correlation coefficients for the model fit and the leave-one-out cross-validation, respectively. Furthermore, we reported the standard errors of fit ($S$) and leave-one-out cross-validation ($S_{cv}$), which were calculated using $N-2$ in the denominator. This statistical procedure and the validation methodology were followed as described in \cite{8}. The scatter plot illustrating this relationship is presented in Fig. 2.  

\begin{figure}[H] 
    \centering
    \includegraphics[width=0.60\linewidth]{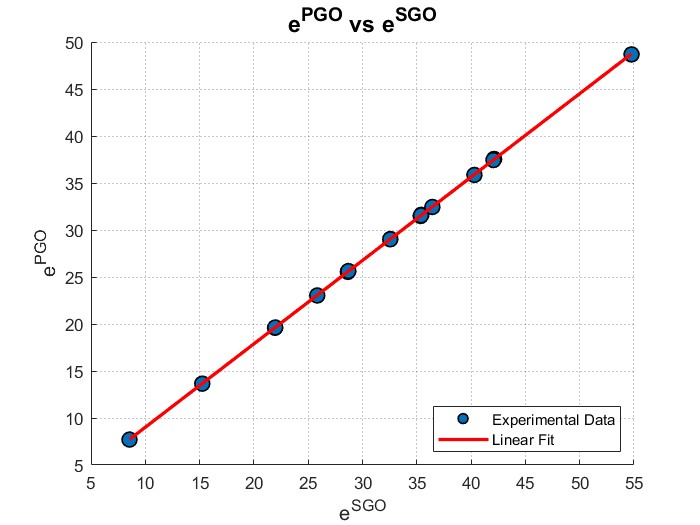} 
    \caption{Scatter plot illustrating the linear correlation between the exponential product-connectivity Gourava index $e^{PGO}$ and the exponential sum-connectivity Gourava index $e^{SGO}$ for a dataset of 30 benzenoid hydrocarbons.}
    \label{fig:scatter_plot}
\end{figure}

Both versions of the exponential connectivity Gourava indices demonstrate a perfect linear intercorrelation ($R = 1.0000$). This indicates that the exponential sum-connectivity Gourava index carries the same structural information as its product-based counterpart; consequently, it is expected to yield equally high-quality structure-property-activity relationship (QSAR/QSPR) models for benzenoid hydrocarbons.

\section{The correlation between exponential sum-connectivity and exponential product-connectivity Gourava indices and the $\pi$-electronic energy of benzenoid hydrocarbons}

In this section, we perform a linear regression analysis between the $\pi$-electronic energy ($E_{\pi}$) and both the exponential sum-connectivity and product-connectivity Gourava indices for the 30 lower benzenoids listed in Table 2. The resulting regression equations and associated correlation parameters are presented in Tables 4 and 5.

\begin{table}[ht]
\centering
\caption{Linear regression model for $E_{\pi}$ vs $e^{SGO}$ and statistical validation parameters.}
\label{tab:regression_expSCG}
\begin{tabular}{ll}
\toprule
\textbf{Parameter} & \textbf{Value} \\
\midrule
Regression equation & $E_{\pi} = (0.8398 \pm 0.0042)\,e^{SGO} + (1.0108 \pm 0.1410) \quad \text{(10)}$ \\
$N$   & 30 \\
$S$ & 0.1912 \\
$R$   & 0.9996 \\
$S_{cv}$ & 0.2219 \\
$R_{cv}$ & 0.9995 \\
\bottomrule
\end{tabular}
\end{table}

\begin{table}[ht]
\centering
\caption{Linear regression model for $E_{\pi}$ vs $e^{PGO}$ and statistical validation parameters.}
\label{tab:regression_expPCG}
\begin{tabular}{ll}
\toprule
\textbf{Parameter} & \textbf{Value} \\
\midrule
Regression equation & $E_{\pi} = (0.9464 \pm 0.0042)\,e^{PGO} + (0.8747 \pm 0.1241) \quad \text{(11)}$ \\
$N$   & 30 \\
$S$ & 0.1675 \\
$R$   & 0.9997 \\
$S_{cv}$ & 0.1945 \\
$R_{cv}$ & 0.9996 \\
\bottomrule
\end{tabular}
\end{table}

The relationships presented in (10) and (11) indicate that the $(e^{SGO})$ and $(e^{PGO})$ indices serve as highly accurate predictors of the $\pi$-electronic energies ($E_{\pi}$) of lower benzenoid structures. However, a comparative analysis of the validation parameters reveals that the exponential product-connectivity Gourava index exhibits superior predictive performance. With a higher correlation coefficient ($R = 0.9997$) and a lower standard error ($S = 0.1675$), the product-based model offers greater statistical stability and accuracy than the exponential sum-connectivity version.

Scatter plots illustrating the relationships between $E_{\pi}$ and the exponential sum-connectivity Gourava index $e^{SGO}$ and product-connectivity Gourava index $e^{PGO}$, respectively, are provided in Fig. 3.

\begin{center}
\begin{figure}[H] 
    \centering
    \includegraphics[width=0.75\linewidth]{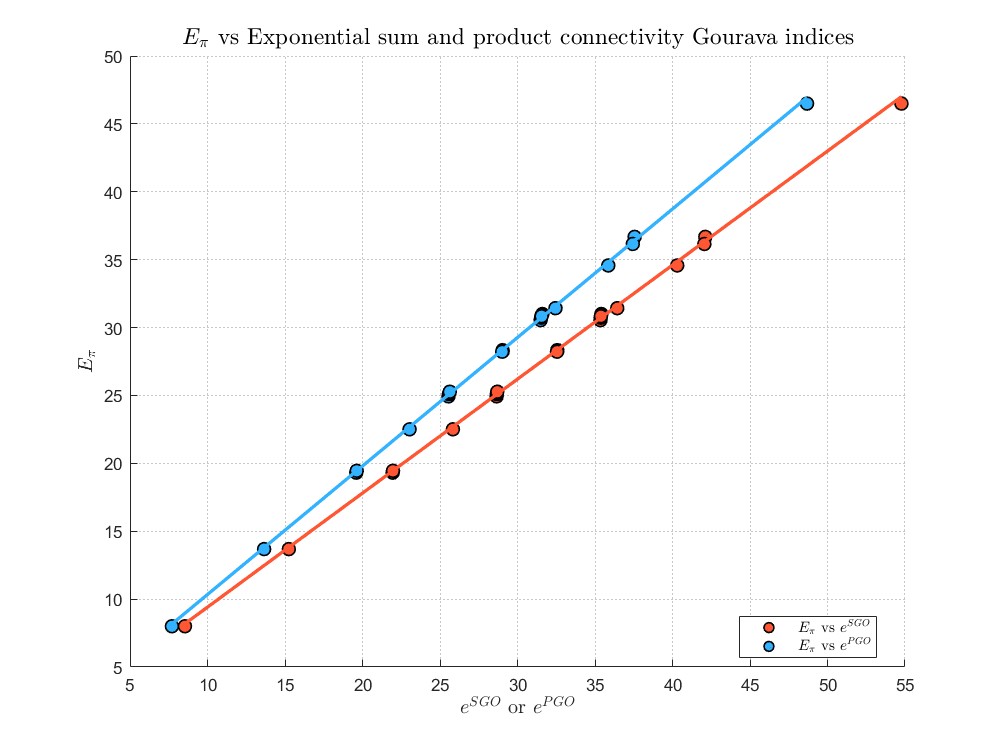} 
    \caption{Scatter plots illustrating the relationships between $E_{\pi}$, $e^{SGO}$, and $e^{PGO}$.}
    \label{fig:scatter_plot}
\end{figure}
\end{center}

As studied in \cite{8}, in addition to relationships through linear regression provided in Tables 4 and 5, we may express $\pi$-electronic energy in the form:
\begin{equation*}
E_{\pi} = \alpha_{1}+ \alpha_{2}e_{22} + \alpha_{3}e_{23} + \alpha_{4}e_{33}.
\end{equation*}
Regression coefficients $\alpha_{1},\alpha_{2},\alpha_{3}$, and $\alpha_{4}$ are obtained by the least-square fit procedure on the set of 30 benzenoid hydrocarbons from Table 2.
\begin{align*}
E_{\pi} = &-0.024(\pm 0.054) + 1.342(\pm 0.006)e_{22} + 1.145(\pm 0.002)e_{23} + 1.061(\pm 0.003)e_{33} \quad \text{(12)}
\end{align*}
It is evident that the uncertainty of the intercept ($\pm 0.054$) significantly exceeds the absolute value of the parameter itself ($-0.024$). Consequently, this term is statistically insignificant and can be omitted from the model.

Substituting (3) into (10) and (4) into (11) yields the following results:
\begin{equation*}
E_{\pi} = 1.011(\pm 0.141) + 1.196(\pm 0.006)e_{22} + 1.135(\pm 0.006)e_{23} + 1.087(\pm 0.005)e_{33}  \quad \text{(13)} 
\end{equation*}
\begin{equation*}
  E_{\pi} = 0.875(\pm 0.124) + 1.215(\pm 0.005)e_{22} + 1.136(\pm 0.005)e_{23} + 1.084(\pm 0.005)e_{33} \quad \text{(14)}
\end{equation*}

\section{Comparison Analysis}
To determine which model is better, we compare how closely the regression coefficients of the ``test" (13) and (14) match the ``optimal" coefficients in (12). This comparison is detailed in Table 6.

\begin{table}[htbp]
    \centering
    \caption{Comparison of regression coefficients for (12), (13), and (14).}
    \label{tab:coeff_comparison}
    \begin{tabular}{lcccc}
        \toprule
        Model & Intercept ($A$) & $e_{22}$ & $e_{23}$ & $e_{33}$ \\
        \midrule
        Eq. (12) Optimal & $-0.024$ & $1.342$ & $1.145$ & $1.061$ \\
        Eq. (13)         & $1.011$  & $1.196$ & $1.135$ & $1.087$ \\
        Eq. (14)         & $0.875$  & $1.215$ & $1.136$ & $1.084$ \\
        \bottomrule
    \end{tabular}
\end{table}

\noindent \textbf{Analysis and Interpretation:} 
One can observe from the error margin of the regression coefficient $A$ in (12) that this parameter is statistically insignificant, as the error ($\pm 0.054$) exceeds the absolute value of the coefficient ($-0.024$). In comparing the test models, both (13) and (14) exhibit positive intercepts, with (14) being slightly closer to the ideal baseline of zero than (13). However, a detailed evaluation of the variable coefficients reveals that both models align exceptionally well with the ``optimal'' coefficients obtained by the least-squares fit in (12). Specifically, for the edge-weight $e_{22}$, the value yielded by (14) ($1.215$) is substantially closer to the ideal value ($1.342$) than that in (13) ($1.196$). For $e_{23}$ and $e_{33}$, both models show nearly identical performance, with (14) maintaining a marginal advantage in proximity to the optimal parameters. This comparative analysis demonstrates that the weighting scheme utilized in the exponential product-connectivity Gourava index (14) captures the relative contributions of different edge types with greater fidelity than the exponential sum-connectivity Gourava index in (13). Consequently, the index represented by (14) emerges as the superior predictor for the $\pi$-electronic energies of this specific set of benzenoid molecules.


\section{Conclusion}
In this study, we have demonstrated that both the exponential sum-connectivity Gourava index ($e^{SGO}(G)$) and the exponential product-connectivity Gourava index ($e^{PGO}(G)$) are closely related and effective molecular descriptors for characterizing the topological properties of molecules. Our comparative regression analysis reveals that both indices provide highly accurate predictions for $\pi$-electronic energies, with correlation coefficients exceeding $0.999$. Specifically, the exponential product-connectivity Gourava index (as represented in (14)) provides a slightly better fit for the relative contributions of various edge-types compared to the exponential sum-connectivity Gourava index (as seen in (13)), as its coefficients align more closely with the optimal least-squares results of (12). These findings suggest that while both variants are robust, the exponential product-connectivity Gourava index may offer superior predictive power for the $\pi$-electronic properties of lower benzenoids. We feel this warrants further investigation into the mathematical properties and broader chemical applications of these exponential sum-connectivity and product-connectivity Gourava indices in quantitative structure-property relationship (QSPR) studies across diverse molecular datasets.

\section*{Funding} No funding is available for this study.
\section*{Author contributions}  H. M. Nagesh: Conceptualization, methodology, and original draft writing. B. Azghar Pasha: Conceptualization. U. Vijaya Chandra Kumar: Methodology. Narahari N.: Validation.
\section*{Data and Software Availability}
The data used to support the work are cited within the text as references. Software used: Python.
\section*{Declarations}
\textbf{Ethical Approval} Not applicable.\\
\textbf{Conflict of interests} The authors declare that they have no known competing financial interests or personal relationships that could have appeared to influence the work reported in this paper.

\end{document}